\documentclass[11pt]{article}
\usepackage[margin=1in]{geometry}
\usepackage[T1]{fontenc}
\usepackage{lmodern}
\usepackage{amsmath,amssymb}
\usepackage{graphicx}
\usepackage{booktabs}
\usepackage{multirow}
\usepackage{xcolor}
\usepackage[numbers,sort&compress]{natbib}
\usepackage[hidelinks]{hyperref}
\usepackage{url}
\newcommand{\syn}{\textsc{[synthetic]}}
\newcommand{\real}{\textsc{[real]}}
\newcommand{\best}{\textsc{[best case]}}
\newcommand{\st}{\,st}

\newif\ifclosedsetpending
\closedsetpendingfalse

\title{Whistle and Wrong Lyrics: Synthetic Queries Are Not Enough\\for Real-World Query-by-Melody}
\author{Luigi Rivolta\\Independent Researcher\\\texttt{luigi@luigirivolta.com}}
\date{October 2026}

\begin{document}
\maketitle

\begin{abstract}
People look up a song by humming it, but also by whistling it or singing it with the wrong words. Published query-by-humming systems and benchmarks focus on hums and sung queries; whistles and wrong-lyrics singing are rarely studied, and no public dataset pairs them with song recordings. We study both query types with a melody-only retrieval system: a key-normalized pitch track (F0) embedded by a 5.2M-parameter contour encoder, with a spectral-peak tracker for whistles. We make three contributions. First, a physics-grounded synthetic query generator that renders whistles (a near-sinusoid with a weak second harmonic, placed in the human whistle register) and nonsense-syllable singing (a source-filter voice) from the vocal melodies of 11,083 songs, then re-tracks each rendering with the tracker a real query would get. Second, a controlled study over three training rounds. Training on these queries raises top-1 retrieval of held-out synthetic queries against an 11,050-song gallery from 0.180 to 0.251 for whistles and from 0.193 to 0.264 for wrong-lyrics singing, at a hum cost of under one point. The gain does not transfer to real whistles: on the MLEnd test performers the best synthetic-only model gets 210 of 339 whistles right at top-1, against 214 for the baseline. Adding real whistles of 6 of the 8 test tunes to training helps those tunes and drops the 2 held-out tunes from 53 to 41 of 90, and a saturated selection metric silently returned the starting weights in the first round. Third, a closed-set real-whistle fine-tune on MLEnd's 8 songs with performer-disjoint splits raises whistle$\to$hum top-1 from 0.631 (E2b zero-shot; clip 95\% CI $[0.581, 0.681]$) to 0.673 with all train performers (CI $[0.622, 0.723]$); this is closed-set on known songs, not open-set generalization. We then test two further generations of synthetic queries. A commercial neural music model conditioned on separated song vocals renders melody-faithful whistles and la-la singing (180 clips); fine-tuning on them leaves real-whistle top-1 at 0.631. An open-source pipeline (YingMusic-Singer-Plus sung queries with partly wrong, mumbled or nonsense lyrics, a DDSP whistle synthesizer, and a melody-error model fitted on real MLEnd training performers) produces about 2,000 queries over 167 songs; no fine-tune beats the baseline on real-whistle top-1 (best 0.619, paired difference $-0.012$, 95\% CI $[-0.038, +0.015]$), and full training on the mixed data costs about 9 points. Across three generations of synthetic data (physics-based, commercial neural, and open-source neural with learned human errors), nothing transferred to real whistles. This points to collecting real queries as the missing step. Code is public; models and data will be released.
\end{abstract}

\section{Introduction}

You probably know the feeling: a song has been stuck in your head all day, you know exactly how it goes, but you cannot remember what it is called, and when you hum it to a friend they look at you as if you have lost your mind. A tune that loops in your head like this is called an \emph{earworm}. Music recognition apps such as Shazam cannot help, because they need the recording itself to be playing, and the whole problem is that it is not. This paper is an attempt at a Shazam for exactly that situation, when all you have is the tune in your head.

Often you do not even know the words. You whistle it, or sing it with whatever words come to mind. The retrieval problem is the same each time: find the recording whose melody matches the query's melody, whatever the timbre, words, key and tempo.

Google's Hum to Search reports that a model trained mostly on sung clips ``worked well enough for people singing, but not for people humming or whistling'' \citep{frank2020hum}. The open query-by-humming (QbH) literature, including CHAD \citep{amatov2023chad} and ByteHum \citep{du2024bytehum}, evaluates hummed and sung queries, not whistles. We found no public dataset of whistled or wrong-lyrics queries paired with the song recordings.

\paragraph{Why whistles are hard.} A whistle is a different sound source from the voice (Section~\ref{sec:signals}): it is close to a pure tone and sits about three octaves above a hum. A vocal pitch model can report it an octave low, and people whistle a compressed version of the tune.

\paragraph{Why wrong lyrics matter.} A spectrogram model can learn to use lyrics and timbre, which a wrong-lyrics query does not share with the recording. A melody-only model cannot use them by design. What stays hard is that made-up words change the voiced/unvoiced pattern and add consonant breaks that a pitch tracker turns into gaps and glitches.

\paragraph{Contributions.}
\begin{enumerate}
\item A physics-grounded synthetic query generator for whistles and nonsense-syllable singing, locked frame by frame to the vocal melody of real songs and re-tracked with the tracker a real query would get (Section~\ref{sec:gen}).
\item A controlled study of synthetic-query training over three rounds (D-040, D-041, D-042). It improves synthetic-query retrieval by about 7 top-1 points across 11k songs but does not transfer to real whistles. It also documents two pitfalls: training on real whistles of the test tunes teaches the tunes, not whistling, and a saturated selection metric returns the starting checkpoint (Sections~\ref{sec:results}--\ref{sec:lessons}).
\item A closed-set real-whistle fine-tune on the 8 songs of MLEnd Hums and Whistles \citep{mlend}: performer-disjoint, E2b zero-shot whistle$\to$hum top-1 0.631 $\to$ 0.673 with all train performers (Section~\ref{sec:closedset}).
\item Melody-conditioned queries from a neural music model (ElevenLabs \texttt{music\_v2\_5} conditioned on separated vocals), with copyright-screen and cost limits, and a negative fine-tuning result on real whistles (Section~\ref{sec:elevenlabs}).
\item Open-source melody-conditioned queries (YingMusic-Singer-Plus singing with wrong, mumbled or nonsense lyrics, a DDSP whistle synthesizer, and a human melody-error model fitted on real performers), with a negative fine-tuning result on real whistles reported with confidence intervals and paired differences (Section~\ref{sec:oss}).
\item An open-source release (Section~\ref{sec:release}).
\end{enumerate}

Taken together, three generations of synthetic queries (physics-based renderings, a commercial neural model, and open-source neural models with learned human errors) improve fit to synthetic queries but none improves retrieval of real whistles. We read this as evidence that real queries, not better synthesis, are the bottleneck.

Throughout, results on generated queries are marked \syn, results on recordings of people are marked \real, and single hand-picked examples are marked \best. Every number comes from a result file in the repository; decision identifiers (D-xxx) refer to entries in the project's decision log.

\section{Related work}

\paragraph{Commercial and large-scale QbH.} Google Hum to Search \citep{frank2020hum} (a blog post, with no published accuracy) embeds spectrograms of queries and recordings with metric learning. Its first training set was mostly sung; synthetic hums were then generated from SPICE pitch \citep{gfeller2020spice}, first as discrete tones and later with an unpublished neural generator. ByteHum \citep{du2024bytehum} trains a CNN on raw audio with weak supervision from source-separated cover songs and unsupervised domain adaptation toward hums, and re-annotates three QbH benchmarks with original recordings. We could not read its full text and report none of its numbers.

\paragraph{CHAD.} \citet{amatov2023chad} collect time-aligned cover fragments with a semi-supervised pipeline plus crowdsourced hums, and train ResNet18 on CREPE \citep{kim2018crepe} F0 activations or CQT with NT-Xent. With CREPE features they report top-10 0.921 on MIR-QBSH against about 2,600 MIDI references and 0.883 on MTG-QBH with 1,886 imposter songs. They note that ``the pure search-by-humming setup requires adding a portion of real humming data''. Our negative result is the whistle analogue of that remark.

\paragraph{Query-by-whistling.} \citet{shen2007whistle} built ``Whistle for Music'', which transcribes whistles to MIDI and uses approximate string matching. \citet{dias2008whistling} compare pitch detectors for whistle transcription. \citet{nilsson2008whistle} report a typical human whistle range of 500--5,000\,Hz. These systems transcribe to notes and match symbols; none learns a joint embedding with real song recordings.

\paragraph{Datasets.} HumTrans \citep{liu2024humtrans}: about 56 hours of hums of 1,000 segments with MIDI labels and official splits. MIR-QBSH \citep{jang2008progressive,mirqbsh}: 4,431 hummed or sung queries of 48 MIDI songs. MLEnd Hums and Whistles \citep{mlend}: hums and whistles of 8 songs from about 200 people, with no song recordings. MTG-QBH \citep{salamon2013tonal}: 118 sung queries; the collection audio is not distributed.

\paragraph{Models and losses.} MERT \citep{li2024mert} is a self-supervised music audio model; we use MERT-v1-95M as a baseline. RMVPE \citep{wei2023rmvpe} estimates vocal pitch from mixtures. CLEWS \citep{serra2025clews} is a segment-level contrastive loss for version matching, which we use for query-to-song windows. Neural singing synthesizers such as SoulX-Singer \citep{soulxsinger2026}, whose melody-control mode takes lyrics and a continuous F0 contour, and Vevo2 \citep{zhang2025vevo2} could replace our simple singer; neither was used here. We use YingMusic-Singer-Plus \citep{yingmusic2026}, which takes a melody-providing clip, a timbre prompt and new lyrics without manual alignment (Section~\ref{sec:oss}), and a small DDSP-style synthesizer \citep{engel2020ddsp} for whistles.

\section{Signals}\label{sec:signals}

All signal physics used in the paper is defined here once and referenced as S1--S8.

\begin{figure}[t]
\centering
\includegraphics[width=0.8\linewidth]{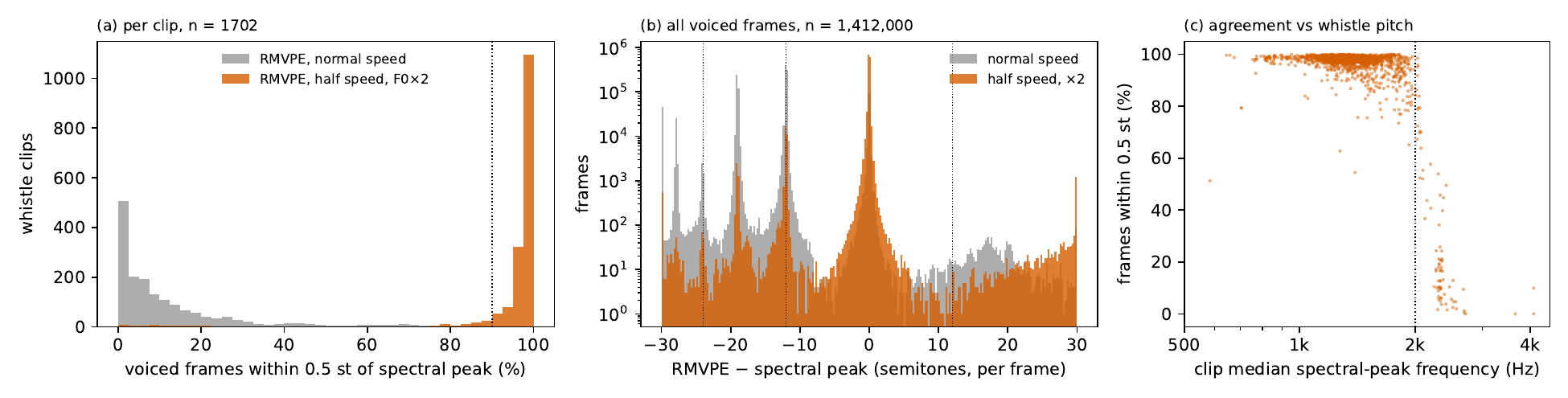}
\caption{Whistle F0 validation on MLEnd (S8): spectral-peak tracker vs.\ half-speed RMVPE. \real}
\label{fig:f0val}
\end{figure}

\begin{description}
\item[S1. F0 and pitch.] A periodic sound has a fundamental frequency $f_0$ and harmonics at $k f_0$. Pitch in semitones is $p = 12\log_2(f_0/440\,\mathrm{Hz}) + 69$. An F0 track is $p$ (or unvoiced) every 10\,ms.
\item[S2. Two sources.] A voice (sung, hummed) is a harmonic source shaped by a vocal-tract filter: many harmonics with resonances (formants). A whistle is a near-sinusoid: on 1,702 MLEnd whistles the median second harmonic is $-45.0$\,dB relative to the fundamental.
\item[S3. Spectral peak.] For a whistle the strongest spectral peak is $f_0$. For a voice it is often a formant-boosted harmonic, so the peak is not $f_0$.
\item[S4. Ranges.] Human whistles typically lie in 500--5,000\,Hz \citep{nilsson2008whistle}. On MLEnd the median hum is 178.7\,Hz and the median whistle about 1.34--1.36\,kHz; the same performer whistles a median $+35.8$\st{} (about three octaves) above their hum. The 5--95\% pitch range of a whistle is 6.5\st{} against 9.7\st{} for hums of the same songs: people compress the tune when they whistle.
\item[S5. Key normalization.] The encoder sees $p$ minus the median voiced pitch of the crop. This makes it exactly invariant to transposition and removes whole-clip octave errors, but not intermittent ones.
\item[S6. Octave error.] RMVPE at normal speed reports a lower octave on a median 55.4\% of a whistle's voiced frames, which makes whistle-to-hum retrieval chance-level (AUC 0.49--0.53). Running it on audio slowed to half speed (``RMVPE-fix'') moves the whistle into its range.
\item[S7. DTW.] Contour pairs are compared with slope-constrained dynamic time warping in re-ranking.
\item[S8. Whistle tracker validation.] Our whistle tracker takes the spectral peak in 200--6,000\,Hz with a 6\,dB tonal-ratio voicing gate. On all 1,702 usable MLEnd whistles it agrees with RMVPE-fix on a median 99.7\% of jointly voiced frames (Figure~\ref{fig:f0val}). RMVPE-fix degrades above about 2\,kHz, and at 0\,dB SNR the peak tracker voices nothing.
\end{description}

\section{Retrieval system}\label{sec:system}

\paragraph{Front ends.} Hums and singing: RMVPE at 10\,ms, voiced where salience $\ge 0.3$, then cleaning (short runs dropped, isolated octave jumps folded, median filter). Whistles: the spectral-peak tracker (S3, S8). Songs: an htdemucs \citep{defossez2021hybrid} vocal stem, then RMVPE, with the mixture as fallback.

\paragraph{Contour encoder.} Input is key-normalized pitch (S5) plus a voicing flag at 20\,ms. A convolutional front end halves the frame rate; a 6-layer pre-norm transformer ($d{=}256$, 4 heads) and a masked attention pool give one L2-normalized 256-d vector (about 5.2M parameters, one tower for queries and songs). The baseline, E2b, was trained with symmetric InfoNCE on HumTrans hum--MIDI pairs and then fine-tuned with the CLEWS loss on cover--original windows from CHAD.

\paragraph{Search.} For the experiments below, each query is scored against up to 24 windows of 8\,s of every song in memory, and a song's score is its best window.

\section{Synthetic whistle and wrong-lyrics queries}\label{sec:gen}

\paragraph{Design principle.} An earlier attempt (D-031) edited contours directly (interval compression, octave folding, gaps). It never passed through a tracker, gave no gain on all 8 MLEnd songs, and cost 4.2 top-1 points on MTG-QBH. We therefore render \emph{audio} and re-track it, so training contours carry the errors a real tracker makes on that kind of sound. Each rendering is locked to a real song's vocal F0, so the positive pair is known exactly.

\paragraph{Source melodies.} The cached vocal-stem F0 of 11,083 open songs (FMA \citep{defferrard2017fma} and chart songs used only as features; evaluation-target titles removed). Four windows per song, each rendered in both styles, gave 86,110 query contours. Songs are split by \texttt{crc32(song\_id) mod 10}: 0 test, 1 validation, otherwise train (68,714 training pairs).

\paragraph{Human errors (both styles).} Before rendering: one note moved by $\pm1$ or $\pm2$\st{} ($p{=}0.5$), one note dropped ($p{=}0.4$), then tempo stretch 0.5--1.9, local warp 0.3, interval scale 0.25, drift 1\st, jitter 0.25\st, voicing dropout 0.6, octave slips 0.15, glides and vibrato.

\paragraph{Whistle.} Register: add 2 or 3 octaves, then shift by whole octaves until the median lies in 500--4,000\,Hz (lower edge S4; upper edge where the peak tracker stayed correct, S8). Tone: a sinusoid at $f_0$ plus a second harmonic drawn uniformly in $-50$ to $-40$\,dB, centered on the MLEnd median (S2). Breath: high-passed noise 30--40\,dB below the tone. Room: one reflection (12--35\,ms, gain 0.04--0.18) plus noise at 18--36\,dB SNR. Re-tracked with the spectral peak (S8).

\paragraph{Wrong-lyrics singing.} Source: 16 harmonics of $f_0$ with a $1/k$ tilt (S2). Filter: three fixed resonances per vowel (a, e, i, o, u), separate settings for nasals and liquids, a short burst for ``d''. Words: syllables of 180--280\,ms drawn at random from \emph{la, na, ma, du, lu, ni, me, lo}; no lyric text is an input, so the words can never be the song's words. Same room. Re-tracked with RMVPE, the tracker real sung queries get.

\paragraph{Renderer check \best.} On one invented tune, the rendered whistle (median 1,558\,Hz) re-tracks with a median absolute error of 0.045\st, and the rendered singing (median 390\,Hz) with 0.106\st. This checks the renderer on one clip; it is not a dataset error rate.

\section{Training rounds and evaluation}\label{sec:setup}

All rounds start from E2b and use one seed. The loss is CLEWS (weight 1.0) between a synthetic query and 4 vocal-melody windows of the same song within $\pm3$\,s (64 songs per batch), plus $0.15\times$ InfoNCE on HumTrans hum--MIDI pairs to keep hums intact. The 0.15 was fixed in advance.

\begin{description}
\item[D-040.] 3,000 steps, LR $3\times10^{-5}$. Selection on HumTrans validation top-1 (query shifted $+7$\st). We report the final checkpoint (step 3,000); Section~\ref{sec:selection} explains why.
\item[D-041.] Selection on the synthetic validation split (mean whistle and wrong-lyrics top-1, with a HumTrans guard), about 4$\times$ longer training, per-epoch query augmentation, and real MLEnd whistle--hum pairs from \emph{training} performers of 6 of the 8 songs. Best checkpoint: step 9,500.
\item[D-042.] D-041 with all MLEnd data removed. Best checkpoint: step 11,250 of 12,000.
\end{description}

\paragraph{Evaluation sets.}
\begin{itemize}
\item \textbf{Synthetic hold-out} \syn: 8,526 queries (4,263 per style) from test-split songs never trained on, against an 11,050-song gallery (training songs stay as distractors).
\item \textbf{Hum baseline} \real: HumTrans test (769 hums) and MIR-QBSH (4,431 queries, 48 MIDI songs, windows anywhere in the song, no distractors), MIDI references.
\item \textbf{MLEnd closed set} \real: Section~\ref{sec:mlend}.
\end{itemize}
Metrics are top-1, top-10 (top-3 for 8-way MLEnd) and mean reciprocal rank (MRR).

\section{Results}\label{sec:results}

\begin{figure}[t]
\centering
\includegraphics[width=\linewidth]{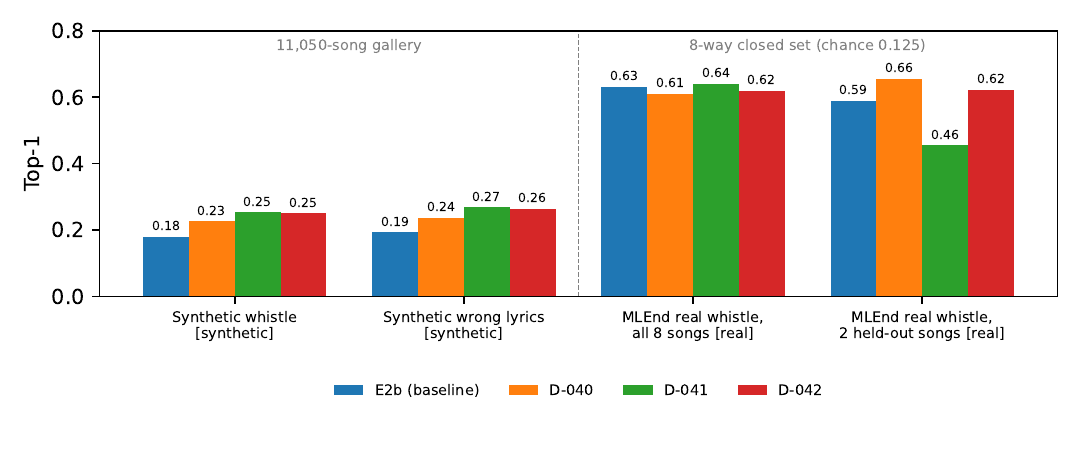}
\caption{Top-1 by model and query set, generated from the committed result files. Left: synthetic hold-out queries \syn. Right: MLEnd real whistles against other people's hums \real. D-041's all-8 bar is not clean (it trained on real whistles of 6 of those songs); its held-out bar is.}
\label{fig:results}
\end{figure}

\subsection{Synthetic queries \syn}

\begin{table}[t]
\centering
\caption{Synthetic hold-out \syn: 4,263 queries per style, 11,050-song gallery. Top-1 / top-10 / MRR. One seed.}
\label{tab:synth}
\small
\begin{tabular}{lccc}
\toprule
Model & Whistle & Wrong lyrics & Both (8,526) \\
\midrule
E2b (baseline) & 0.180 / 0.291 / 0.220 & 0.193 / 0.314 / 0.235 & 0.186 / 0.303 / 0.227 \\
D-040 (step 3,000) & 0.226 / 0.342 / 0.267 & 0.238 / 0.365 / 0.283 & 0.232 / 0.354 / 0.275 \\
D-041 best (9,500) & 0.253 / 0.378 / 0.297 & 0.269 / 0.401 / 0.314 & 0.261 / 0.389 / 0.305 \\
D-042 best (11,250) & 0.251 / 0.374 / 0.295 & 0.264 / 0.399 / 0.312 & 0.258 / 0.387 / 0.303 \\
\bottomrule
\end{tabular}
\end{table}

Table~\ref{tab:synth} shows a clear, steady gain: D-042 adds 0.071 to 0.072 top-1 over E2b in both styles. With 4,263 queries per style, the binomial standard error of a top-1 near 0.2 is about 0.006, so the gain is far outside query-sampling noise (seed variance is not measured). Removing the MLEnd pairs (D-041 $\to$ D-042) costs at most 0.005, so D-041's synthetic gain came from selection, length and augmentation. These queries come from the same generator used in training, on new melodies: the table measures fit to the generator, not robustness to people.

\subsection{Hum baseline \real}

\begin{table}[t]
\centering
\caption{Real hums against MIDI \real. Top-1 / top-10.}
\label{tab:hum}
\small
\begin{tabular}{lcccc}
\toprule
Set & E2b & D-040 & D-041 best & D-042 best \\
\midrule
HumTrans test (769) & 0.996 / 1.000 & 0.991 / 1.000 & 0.993 / 1.000 & 0.992 / 1.000 \\
HumTrans test, $+7$\st & 0.996 / 1.000 & 0.995 / 1.000 & 0.995 / 1.000 & 0.993 / 1.000 \\
MIR-QBSH 48 (4,431) & 0.955 / 0.993 & 0.944 / 0.991 & 0.949 / 0.991 & 0.950 / 0.992 \\
\bottomrule
\end{tabular}
\end{table}

The small hum term keeps hums nearly intact (Table~\ref{tab:hum}). D-042 costs 0.004 HumTrans top-1 (763 vs.\ 766 of 769) and 0.005 MIR-QBSH (4,209 vs.\ 4,232 of 4,431).

\subsection{Closed-set real whistles on MLEnd \real}\label{sec:mlend}

\paragraph{Protocol.} MLEnd has no song recordings, so we cannot score a real whistle against the song itself. Instead, each test performer's whistle (spectral-peak tracker) is scored against the centroid embedding of \emph{other people's} hums of each of the 8 songs; the query's own performer is excluded. This is an 8-way closed set (chance top-1 0.125) over 339 whistles from 26 test whistlers. Hakuna and Potter are the two songs held out from every MLEnd training pair. Neither E2b, D-040 nor D-042 trained on MLEnd, so every row is a clean test for them; for D-041 only the held-out row is clean.

\begin{table}[t]
\centering
\caption{MLEnd test whistles $\to$ other people's hums, 8-way closed set \real. Top-1 / top-3 / MRR, with correct top-1 counts in parentheses.}
\label{tab:mlend}
\small
\setlength{\tabcolsep}{4pt}
\begin{tabular}{lccc}
\toprule
Model & Hakuna + Potter (90) & Other 6 songs (249) & All 8 (339) \\
\midrule
E2b & 0.589 / 0.711 / 0.698 (53) & 0.647 / 0.791 / 0.742 (161) & 0.631 / 0.770 / 0.731 (214) \\
D-040 & 0.656 / 0.789 / 0.741 (59) & 0.594 / 0.779 / 0.720 (148) & 0.611 / 0.782 / 0.725 (207) \\
D-041 best$^\dagger$ & 0.456 / 0.700 / 0.620 (41) & 0.707 / 0.863 / 0.803 (176) & 0.640 / 0.820 / 0.754 (217) \\
D-042 best & 0.622 / 0.722 / 0.709 (56) & 0.618 / 0.763 / 0.724 (154) & 0.619 / 0.752 / 0.720 (210) \\
\bottomrule
\multicolumn{4}{l}{\footnotesize $^\dagger$Trained on real whistles of the other 6 songs: only the Hakuna + Potter column is clean.}
\end{tabular}
\end{table}

\paragraph{No transfer.} On the full clean test, D-042 gets 210 of 339 whistles right against 214 for E2b (Table~\ref{tab:mlend}). Its $+3$ of 90 on the held-out pair is inside the binomial standard error (about 0.05) and is offset by $-7$ of 249 on the other six. D-040 shows the same pattern. Three rounds of synthetic training raised synthetic whistle top-1 by 0.07 and did not move real whistles beyond noise.

\paragraph{Real whistles of the test tunes teach the tunes.} D-041, which also trained on real whistles of 6 of the 8 songs from other performers, gains 0.060 top-1 on those 6 songs but falls to 41 of 90 on the 2 it never saw, against 53 for E2b and 59 for D-040. Removing the pairs (D-042) recovers 56 of 90. The same pattern appeared in an earlier experiment (D-015/D-016: held-out songs 0.630 $\to$ 0.456). In a closed set this small, real pairs teach song identity rather than the whistle-to-hum mapping, and an all-8 number that mixes trained and untrained songs (0.640 for D-041) should not be read as a whistle gain.

\paragraph{Tracker matters more than training.} With the same E2b weights, whistle top-1 over all 8 songs is 0.631 with the spectral-peak tracker, 0.584 with RMVPE-fix and 0.159 with plain RMVPE, consistent with S6.

\subsection{Closed-set real-whistle training}\label{sec:closedset}

Following the scoped design, E2b is fine-tuned on real MLEnd train-performer whistle$\to$hum and hum$\to$hum pairs and tested on held-out performers within the same 8 songs (D-044; CPU run). Selection uses val-performer whistle$\to$hums top-1. Scoring matches Section~\ref{sec:mlend}: spectral-peak whistle against the centroid of other people's hums, query performer left out (339 test whistles, 26 people; chance top-1 0.125). Nested subsets of 10/25/50/100\% of the 136 train performers give a low-data curve (one seed). This is a closed-set result by construction: it says how well real whistle training works when the song set is fixed and known, not how it generalizes to new songs.

\begin{table}[t]
\centering
\caption{Closed-set real-whistle training on MLEnd's 8 songs \real (D-044). Whistle $\to$ other people's hums, 8-way; test performers disjoint from training. Top-1 shows clip 95\% CI; MRR is mean $1/(\mathrm{rank}+1)$ from the same ranks. Source: \texttt{docs/paper/results/d044/results.json}.}
\label{tab:closedset}
\small
\setlength{\tabcolsep}{4pt}
\begin{tabular}{lcccc}
\toprule
Model & Train performers & Top-1 [clip 95\% CI] & Top-3 & MRR \\
\midrule
E2b zero-shot & 0 & 0.631 [0.581, 0.681] & 0.770 & 0.731 \\
FT 10\% & 14 & 0.634 [0.584, 0.684] & 0.776 & 0.735 \\
FT 25\% & 34 & 0.625 [0.575, 0.678] & 0.794 & 0.738 \\
FT 50\% & 68 & 0.655 [0.605, 0.705] & 0.811 & 0.757 \\
FT 100\% & 136 & 0.673 [0.622, 0.723] & 0.820 & 0.768 \\
\bottomrule
\end{tabular}
\end{table}

Full training reaches 228 of 339 top-1 (0.673) against 214 for E2b (0.631), and top-3 0.820 against 0.770. Clip CIs overlap; performer-bootstrap CIs are wide (E2b $[0.500, 0.753]$, full FT $[0.536, 0.801]$). The curve is roughly flat at 10--25\% and rises at 50--100\%. These weights are not adopted for open retrieval: the live model stays E2b.

\subsection{MERT baseline \real}

A fine-tuned MERT-v1-95M encoder (8,000 further steps on 13,080 HumTrans hums) reaches HumTrans test top-1 / top-10 0.947 / 0.999 but only 0.169 / 0.602 on MIR-QBSH (48 songs, first 10\,s of each reference). The contour encoder E2b scores 0.996 / 1.000 and 0.949 / 0.990 on the same sets. MERT learns HumTrans but does not transfer to MIR-QBSH; the F0 encoder does, which motivates the contour design. The two evaluation scripts differ, so the HumTrans rows are close but not guaranteed identical protocols.

\section{Neural music-model queries (ElevenLabs)}\label{sec:elevenlabs}
As a second source of synthetic queries we tried a commercial neural music model (D-045--D-047). Prompting alone fails: whistles and ``la-la'' singing asked for by song name sound realistic but do not follow the melody (0 of 12 prompt-only clips passed a DTW melody check against the reference F0) \syn. Conditioning works. We separate the vocal of a song section with htdemucs, upload it to the ElevenLabs Music API, and generate a 20\,s clip from \texttt{music\_v2\_5} with that upload as \texttt{conditioning\_ref}. Whistles use condition strength \emph{xhigh} and a solo-whistling prompt. La-la uses \emph{medium} with amateur female or young-male voice prompts; \emph{low} lost the tune by ear. Each song gets up to three 20\,s windows with the most vocal activity, and each accepted section gets one whistle and one la-la clip. The clips follow the tune by ear. With E2b as a judge on a closed set of 10 songs, 8 of 10 la-la clips and 3 of 10 whistles rank the right song first (chance 0.1) \syn. Against the 1,051-song judge set (\texttt{youtube\_charts\_v1} + \texttt{youtube\_v1}), top-1 rates are lower (Table~\ref{tab:elevenlabs}).

\begin{table}[t]
\centering
\caption{ElevenLabs conditioned queries \syn (D-045, D-046) and fine-tuning on them, scored on real whistles \real (D-047). Judge: E2b, top-1 against 1,051 songs. Uploads rejected by the provider's copyright screen are still billed.}
\label{tab:elevenlabs}
\small
\setlength{\tabcolsep}{4pt}
\begin{tabular}{lcc}
\toprule
 & Phase A (32 classics) & Phase B (79 chart songs) \\
\midrule
Sections accepted / blocked & 37 / 55 & 54 / 183 \\
Clips (whistle + la-la) & 74 & 106 \\
La-la top-1 / top-10 & 6/37 / 11/37 & 17/53 / 22/53 \\
Whistle top-1 / top-10 & 3/37 / 11/37 & 7/53 / 11/53 \\
\midrule
\multicolumn{3}{l}{MLEnd test whistles $\to$ hums \real: E2b 0.631 [0.581, 0.681] (339), 0.589 [0.489, 0.689] (held-out 90);} \\
\multicolumn{3}{l}{D-047 fine-tune: identical (best checkpoint = step 0).} \\
\bottomrule
\end{tabular}
\end{table}

\paragraph{Scale limits.} The provider screens uploads for copyrighted material. In total, 91 of 329 sections were accepted and 238 blocked (72\%). Blocked uploads were billed at half price. About 106k credits (roughly \$22) bought 180 clips: about \$0.12 per clip and about \$0.67 per clip that E2b ranks first. Only 32 songs had audio held locally for phase A, and phase B was stopped after 79 songs.

\paragraph{Fine-tuning result \real.} D-047 fine-tunes E2b on the CPU for 1,500 steps. Each batch is half ElevenLabs pairs (166 clips from 48 songs, excluding songs that may overlap MLEnd's tunes) and half the 68,714 synthetic pairs of Section~\ref{sec:gen}. The checkpoint is selected on MLEnd validation performers. Validation top-1 never exceeded its step-0 value (0.794; 0.710--0.745 afterwards), so the selected checkpoint is E2b itself. On test performers it scores whistle$\to$hum top-1 0.631 [0.581, 0.681] on all 339 and 0.589 [0.489, 0.689] on the 90 held-out-tune whistles, identical to E2b.

\paragraph{Why it did not help.} Three reasons. (i) Scale: 166 clips against about 69k synthetic pairs. (ii) Realism: the clips copy the timing and intonation of the original vocal closely, so they lack the tempo drift, wrong intervals and dropped notes of real amateurs. (iii) Coverage: few songs. We report this as a new way to make queries plus a negative result. Melody-faithful neural renderings do not close the synthetic-to-real gap on their own, which again points to real queries as the bottleneck. Section~\ref{sec:oss} tests the first two of these with open-source models: more songs, no copyright screen, and learned human errors.

\section{Open-source melody-conditioned queries}\label{sec:oss}
The ElevenLabs route was limited by cost, the copyright screen and the lack of human errors. We therefore built a third generation of synthetic queries with open models we can run on our own GPU, and added a model of human melody errors fitted on real performers (D-048, D-049/D-051). All clips are synthetic \syn; all scores below are on real MLEnd whistles \real.

\paragraph{Sung queries.} We use YingMusic-Singer-Plus \citep{yingmusic2026} (code and weights at \url{https://github.com/ASLP-lab/YingMusic-Singer-Plus}, CC BY 4.0; the VAE is under the Stability AI Community License), which sings given lyrics along a melody-providing clip in the voice of a timbre prompt. Melody guide: RMVPE F0 of the separated vocal of a song section, with octave jumps folded and gaps filled (gaps up to 0.3\,s bridged by interpolation, longer gaps held at the previous pitch), rendered as a five-harmonic tone with light vibrato. This keeps the sung query continuous along the tune instead of stopping where the original singer paused. Lyrics come from an English ASR transcript of the section and take one of three forms: \emph{partial-real} (each word replaced with probability 0.3 by a word from another song), \emph{mumbled} (words turned into ``mm''/``hmm'', truncated, or replaced by syllables), and \emph{mixed syllables} (\emph{da, na, la, ooh, doo, \ldots}). Voice prompts are 5--8\,s English vocal snippets from \emph{other} songs (plus two examples shipped with the model), so the timbre never matches the target song.

\paragraph{Whistles.} A DDSP-style synthesizer \citep{engel2020ddsp} (a GRU predicting the amplitude of 8 harmonics plus 48 noise bands from F0 and loudness, trained with a multi-scale spectral loss) was trained on 87 ElevenLabs whistles from Section~\ref{sec:elevenlabs}. It is driven by the same gap-filled guide, shifted by whole octaves into the register of those whistles. Each clip is re-tracked with the front end a real query would get (Section~\ref{sec:system}).

\paragraph{Human melody errors.} For D-049 arm C we fitted per-clip error statistics on 718 hums and whistles of MLEnd \emph{training} performers of 6 songs (Hakuna and Potter excluded), by DTW against a reference hum of the same song: global pitch drift (slope), note-level interval error, frame jitter, key offset, mean tempo ratio and local tempo variability, and drop/insert rates of the alignment. Before rendering, a guide is perturbed with the parameters of one randomly drawn real clip (some scaled down or clipped): a smooth local tempo warp, piecewise-constant interval offsets per note, drift, jitter and a key offset. The fitted drop rates are recorded but not applied as explicit note deletions; dropped notes only enter through the voicing-dropout query augmentation used in all arms.

\paragraph{Training.} D-048 (CPU): 948 usable pairs (639 sung, 309 whistle; 105 songs, from 1,316 generated clips), mixed 50/50 per batch with the 68,714 synthetic pairs of Section~\ref{sec:gen}, LR $3\times10^{-5}$, 1,500 steps. D-049 (one A10 GPU) adds a new batch: 698 of 710 sung jobs returned audio. The pooled set has 1,994 pairs over 167 songs (1,326 sung, 668 whistles; 349 with human errors). Arm A mixes the 1,645 unperturbed pairs 50/50 with Section~\ref{sec:gen}'s pairs (LR $10^{-5}$). Arm B trains on open-source pairs only, with LR $3\times10^{-6}$ and the first 75\% of parameter tensors frozen (63 of 85). Arm C is A plus the 349 perturbed pairs. All runs start from E2b, use one seed and 1,500 steps, and select the checkpoint on MLEnd \emph{validation} performers. Hakuna and Potter are absent from the open-source songs. Two MLEnd tunes (Purple Rain, The Greatest Showman) were among the D-049 source songs, so only the Hakuna + Potter column is fully held out.

\begin{table}[t]
\centering
\caption{Open-source synthetic queries, scored on MLEnd test whistles $\to$ other people's hums \real{} (8-way; D-044 protocol). Top-1 with clip 95\% bootstrap CI; $\Delta$ is the paired bootstrap top-1 difference against E2b on the same clips. ``Best'' = checkpoint selected on MLEnd validation performers; ``last'' = step 1,500. P90 is the melody-filtered subset (Section~\ref{sec:oss}, D-050). Sources: \texttt{docs/paper/results/d048}, \texttt{d049}, \texttt{d050}.}
\label{tab:oss}
\resizebox{\linewidth}{!}{%
\begin{tabular}{lcccccc}
\toprule
 & \multicolumn{2}{c}{All 8 songs (339)} & \multicolumn{3}{c}{Hakuna + Potter, held out (90)} & P90 (304) \\
\cmidrule(lr){2-3}\cmidrule(lr){4-6}\cmidrule(lr){7-7}
Model & Top-1 [CI] & $\Delta$ [CI] & Top-1 [CI] & $\Delta$ [CI] & Top-3 & Top-1 [CI] \\
\midrule
E2b & 0.631 [0.581, 0.681] & -- & 0.589 [0.489, 0.689] & -- & 0.711 & 0.648 [0.592, 0.701] \\
D-048 best (= step 0) & \multicolumn{6}{c}{identical to E2b} \\
D-048 last & 0.546 [0.496, 0.599] & n/a & 0.533 [0.433, 0.633] & n/a & 0.756 & 0.549 [0.493, 0.605] \\
\midrule
D-049 A best (= step 0) & \multicolumn{6}{c}{identical to E2b} \\
D-049 A last & 0.546 [0.493, 0.599] & $-0.086$ [$-0.130$, $-0.044$] & 0.511 [0.411, 0.611] & $-0.078$ [$-0.156$, $0.000$] & 0.733 & 0.553 [0.497, 0.609] \\
D-049 B best (step 750) & 0.619 [0.569, 0.670] & $-0.012$ [$-0.038$, $+0.015$] & 0.589 [0.489, 0.689] & $0.000$ [$-0.044$, $+0.044$] & 0.789 & 0.635 [0.582, 0.688] \\
D-049 B last & 0.593 [0.543, 0.646] & $-0.038$ [$-0.074$, $-0.006$] & 0.611 [0.511, 0.711] & $+0.022$ [$-0.033$, $+0.089$] & 0.811 & 0.602 [0.546, 0.658] \\
D-049 C best (= step 0) & \multicolumn{6}{c}{identical to E2b} \\
D-049 C last & 0.543 [0.490, 0.596] & $-0.088$ [$-0.133$, $-0.044$] & 0.511 [0.411, 0.611] & $-0.078$ [$-0.167$, $+0.011$] & 0.756 & 0.549 [0.493, 0.605] \\
\bottomrule
\end{tabular}}
\end{table}

\paragraph{Results \real.} No run beats E2b on top-1, and no paired top-1 difference has a CI above zero (Table~\ref{tab:oss}). In D-048, and in arms A and C, MLEnd validation top-1 never exceeded its step-0 value (0.794), so the selected checkpoint is E2b itself; the final checkpoints lose about 9 points on all 339 whistles ($-0.086$ and $-0.088$, CIs excluding zero). Adding the learned human errors (C vs.\ A) did not help. Arm B, which changes the model least, is the closest: its selected checkpoint (validation 0.797 at step 750) is a statistical tie on all 339 ($-0.012$, CI $[-0.038, +0.015]$), and its final checkpoint is $+0.022$ on the 90 held-out whistles (CI $[-0.033, +0.089]$) but significantly worse on all 339 ($-0.038$). The one possibly positive signal is secondary: B's held-out top-3 rises from 0.711 to 0.789 (best) and 0.811 (last), and from 0.732 to 0.805 and 0.829 on the 82 held-out whistles that pass the melody filter. It was not pre-registered or significance-tested, and we do not count it as a result.

\paragraph{Robustness: melody-filtered subset (D-050).} One reading of the low absolute numbers is that many MLEnd whistles do not carry the tune. We fixed a filter before scoring, independent of any model: a clip is kept if its whistle tracker voices at least 25\% of frames and its key-invariant subsequence-DTW cost against 40 hums of the same song by \emph{training} performers is below the 90th percentile of the same cost for \emph{validation} performers' hums. It keeps 304 of 339 test whistles (82 of the 90 held-out), dropping mostly Rain, StarWars and Potter clips. E2b rises only from 0.631 to 0.648 [0.592, 0.701], inside the CI, and the ordering of models is unchanged (Table~\ref{tab:oss}, last column; P90 paired differences for B best and last are $-0.013$ $[-0.039, +0.013]$ and $-0.046$ $[-0.082, -0.010]$). So most of E2b's errors are not caused by tune-less clips.

\paragraph{A harsh benchmark.} MLEnd whistles come from about 200 amateur participants, not trained singers or whistlers. They compress the tune (S4), drift and change tempo, and the references are other amateurs' hums, not the song. Absolute top-1 near 0.63 on an 8-way set should be read with that in mind; the comparisons in this paper are all paired on the same clips.

\paragraph{Reading.} The open-source pipeline removes the scale and copyright limits of Section~\ref{sec:elevenlabs} (about 2,000 queries over 167 songs) and adds errors measured on real people, yet the outcome is the same as for the physics-based renderer and the commercial model. In these runs, fine-tuning on synthetic whistles and singing moved the encoder away from what works on real whistles, and the run that changed it least (B) lost least. We did not measure the synthetic-to-real gap directly, so we cannot say which property of real queries is missing, only that three different generators did not supply it.

\section{Lessons}\label{sec:lessons}

\subsection{The checkpoint-selection pitfall}\label{sec:selection}
D-040 selected checkpoints on HumTrans validation top-1 with a $+7$\st{} shift. That metric was 0.9948 at step 0 (the E2b weights) and never exceeded it, drifting to 0.9922. The first run early-stopped at step 500 with ``best step 0''; the second run, with patience off, also selected step 0, so its ``best'' checkpoint was the baseline and every number in its result file equals E2b's exactly. We report the fixed-schedule final checkpoint instead. The fix (D-041) selects on the synthetic validation split with HumTrans as a guard; that metric does move, from 0.288 at step 0 to 0.374 at the best step in both D-041 and D-042, with the guard holding at every validation (0.9908--0.9948). The rule works, but the metric it optimizes is synthetic, so it cannot see the real-whistle behavior in Table~\ref{tab:mlend}. A selection metric must be able to move on what is being trained \emph{and} track the target domain.

\subsection{Other results that shaped the method}
\begin{itemize}
\item Hum fine-tuning with positives built from each song's own distorted F0 (D-039) gave no real gain: CHAD validation top-1 0.4359 at best vs.\ 0.4341 at step 0.
\item Rendering and re-tracking audio (D-040) costs hums far less than editing contours (D-031: $-4.2$ MTG-QBH top-1).
\item The largest real-hum gains in this project came from matching (multi-window voting and re-ranking), not from training.
\end{itemize}

\section{Limitations}\label{sec:limits}
\begin{enumerate}
\item \textbf{No real whistle or wrong-lyrics query is scored against the actual song recording.} MLEnd ships whistles but not the songs, and MTG-QBH ships sung queries but not the recordings. Our real-whistle evidence is whistle $\to$ other people's hums.
\item \textbf{No real wrong-lyrics test.} The wrong-lyrics results are synthetic only.
\item \textbf{MLEnd is small and harsh:} 8 songs and 26 test whistlers, all amateurs; the held-out pair is 2 songs, 90 whistles. Per-song effects dominate. D-049's training songs include 2 of the 8 MLEnd tunes (synthetic queries only), so only its Hakuna + Potter numbers are fully held out.
\item \textbf{Synthetic-to-real gap not measured:} the physics-based renderer is simple (one-reflection room, fixed formants, 8 syllables), but the two neural generators are not, and they failed too. We have not measured which property of real queries the generators miss. The human-error model does not apply explicit note deletions.
\item \textbf{One seed} per round; differences of one or two points are not resolved.
\item \textbf{Song side:} melody comes from the vocal stem, so instrumental hooks are unreachable.
\item \textbf{Whistles above about 2\,kHz} and very noisy whistles are tracked poorly (S8).
\item \textbf{No direct comparison} with Google (no published numbers) or ByteHum (full text not read).
\end{enumerate}

\section{Release}\label{sec:release}
Code, configuration files, decision log and every result file cited here are on GitHub at \url{https://github.com/luiginotmario/hum2song}. Model checkpoints (E2b, D-040 to D-042, MERT runs) and the search index will be released on Hugging Face as \texttt{luigir5/hum2song-models}; derived data (melody features, song catalog, result files) will be released as \texttt{luigir5/hum2song-data}. Both repositories exist and are private at the time of writing; they are to be released.

\section{Conclusion and future work}
A melody-only contour encoder with a whistle-specific tracker is a sound base for whistle and wrong-lyrics search: lyrics and timbre cannot mislead it, and key is removed exactly. Rendering whistles and nonsense-syllable singing from real song melodies and re-tracking them improves retrieval of such queries on new songs by about 7 top-1 points at under one point of hum cost. It does not improve retrieval of real whistles in the open sense, and real whistles of the test tunes teach the tunes rather than whistling. On a closed set of the same 8 MLEnd songs with held-out performers, fine-tuning on real whistles raises top-1 from 0.631 to 0.673 (Table~\ref{tab:closedset}); that gain is not a claim about new songs. Melody-conditioned neural renderings from a commercial model (Section~\ref{sec:elevenlabs}) follow the tune but did not move real-whistle retrieval (0.631 top-1, unchanged). Open-source neural sung and whistled queries over 167 songs, with a melody-error model fitted on real performers (Section~\ref{sec:oss}), did not either: the best run ties E2b (paired top-1 difference $-0.012$ $[-0.038, +0.015]$) and full training on the mixed data loses about 9 points. Three generations of synthetic data (physics-based, commercial neural, open-source neural with learned human errors) failed to transfer to real whistles on a benchmark of amateur performers. As CHAD found for hums, synthetic or proxy data is not enough for open-set transfer: what is missing is real whistled and wrong-lyrics queries paired with recordings. Next steps are (i) collecting such queries for songs whose audio we can hold, (ii) selecting checkpoints on a real-query validation set, (iii) measuring the synthetic-to-real gap before training rather than after, and (iv) more seeds.

\bibliographystyle{plainnat}
\bibliography{refs}

\end{document}